\documentclass[conference]{IEEEtran}
\IEEEoverridecommandlockouts

\usepackage{cite}
\usepackage{amsmath,amssymb,amsfonts}
\usepackage{algorithmic}
\usepackage{graphicx}
\usepackage{textcomp}
\usepackage{xcolor}
\usepackage{gensymb}

\def\BibTeX{{\rm B\kern-.05em{\sc i\kern-.025em b}\kern-.08em
    T\kern-.1667em\lower.7ex\hbox{E}\kern-.125emX}}
    
\begin{document}

\title{Deep Learning-Based Beamforming Design Using Target Beam Patterns}

\author{\IEEEauthorblockN{Hongpu Zhang$^{1}$, Shu Sun$^{1}$, Hangsong Yan$^{2}$, Jianhua Mo$^{1}$}
    \IEEEauthorblockA{$^1$ School of Information Science and Electronic Engineering, Shanghai Jiao Tong University, Shanghai, China}
    \IEEEauthorblockA{$^2$ Hangzhou Institute of Technology, Xidian University, Hangzhou, Zhejiang, China}
    \IEEEauthorblockA{Corresponding author: Shu Sun (Email: shusun@sjtu.edu.cn)}
    \thanks{This work is supported by the National Natural Science Foundation of China under Grants 62271310 and 62431014.}
    }

\maketitle 

\begin{abstract}
This paper proposes a deep learning-based beamforming design framework that directly maps a target beam pattern to optimal beamforming vectors across multiple antenna array architectures, including digital, analog, and hybrid beamforming. The proposed method employs a lightweight encoder-decoder network where the encoder compresses the complex beam pattern into a low-dimensional feature vector and the decoder reconstructs the beamforming vector while satisfying hardware constraints. To address training challenges under diverse and limited channel station information (CSI) conditions, a two-stage training process is introduced, which consists of an offline pre-training for robust feature extraction using an auxiliary module, followed by online training of the decoder with a composite loss function that ensures alignment between the synthesized and target beam patterns in terms of the main lobe shape and side lobe suppression. Simulation results based on NYUSIM-generated channels show that the proposed method can achieve spectral efficiency close to that of fully digital beamforming under limited CSI and outperforms representative existing methods.
\end{abstract}


\section{Introduction}
Massive multiple-input multiple-output (MIMO) offers transformative advantages in wireless communication by leveraging a large array of antennas at base stations (BSs), dramatically enhancing spectral efficiency \cite{BG1}. Advanced beamforming techniques are widely used in massive MIMO, which allow for precise signal directionality, improving signal strength, reducing interference, and extending coverage\cite{BG, MUBEAM}. However, the difficulty in obtaining accurate channel state information (CSI) due to large-scale antenna arrays and the high operating frequency poses challenges to the system design, as conventional beamforming methods heavily depend on precise CSI to maintain the theoretical performance gains. 

Existing beamforming architectures mainly include digital beamforming (DBF), analog beamforming (ABF), and hybrid beamforming (HBF). HBF achieves a balance between hardware cost and spectrum efficiency by jointly optimizing analog and digital domain beamforming matrices. HBF methods include iterative optimization algorithms such as orthogonal matching pursuit (OMP)\cite{OMP}, manifold optimization based alternating minimization (MO-AltMin)\cite{MO}, as well as methods based on deep learning (DL)\cite{DLBF}. For ABF, beam scanning and selection are usually performed by using predefined codebooks\cite{Codebook}. However, optimal DBF and HBF based on optimization or DL require accurate CSI (i.e., accurate channel matrix) to design the beamforming matrix, causing high overhead in channel estimation. Codebook-based methods suffer from quantization errors and require lengthy beam training, while array synthesis is crucial for MIMO beamforming, using beam patterns to derive beamforming vectors. Traditional array synthesis methodologies, including analytical techniques and numerical optimization approaches\cite{GA,ConvP}, are effective in small-scale arrays, but their computational complexity rises sharply as the number of array elements increases, making them unsuitable for large-scale MIMO arrays. Existing DL-based array synthesis methods \cite{AEArraySyn,AE2D} reduce complexity but lack generalization, requiring retraining for new antenna configurations or scenarios. Meanwhile, most of the research focuses on DBF and ABF, but few studies on array synthesis under HBF.

In order to reduce the CSI acquisition overhead in beamforming, we consider using partial CSI for beamforming across diverse antenna array architectural configurations. Specifically, during beamforming, the channel vector between the BS and user equipment (UE) is not available. Instead, only partial information such as the UE’s angular location and beam shape characteristics (e.g., the main lobe direction, shape, and side lobe level) is known. However, such information cannot be directly utilized for the beamforming design based on spectral efficiency maximization. Here we consider using normalized amplitude beam pattern of the channel as the partial CSI, which can be obtained through low-overhead techniques to estimate angles-of-departure (AoDs) and path gain amplitudes\cite{PathLossEst}, fast ray-tracing on 3D maps\cite{RayTracing}, or multi-modal-sensing based user localization. A more detailed study of beam pattern acquisition will be discussed in future work. Moreover, customized beam patterns are demanded in scenarios such as directional broadcasting\cite{modelanalysis} and flat-top coverage. This motivates beamforming approaches that support flexible pattern control while reducing CSI estimation overhead. In this paper, we propose a DL-based beamforming method for all three beamforming architectures using a target beam pattern, which consists of an encoder and a decoder module. Specifically, the encoder extracts low-dimensional features from the complex beam pattern, and then the decoder uses these features to perform beamforming design under each architecture. A two-stage training process is proposed to effectively train the deep neural network (DNN), with pre-training of the encoder and online training for the decoder. Simulation results show the effectiveness of the proposed method in large antenna array synthesis.

\section{System Model and Problem Formulation}
\subsection{System Model}
In this paper, we consider a single-user MIMO communication system where a BS is equipped with an $N_t$-element uniform rectangular array (URA), serving a single-antenna UE. If a multiple-user system is considered, the beamforming design can be derived similarly. The primary focus of this study is on efficient far-field beamforming \cite{FARNEAR} at the BS using a highly limited amount of CSI, specifically the target beam pattern at the BS. Our goal is to determine the optimal beamforming vector $\mathbf{f}$ given the target beam pattern.

Three different beamforming architectures are considered. For DBF, the beamforming vector is indicated as $\mathbf{f}_{fd} \in \mathbb{C}^{N_t \times 1}$, where its elements are unconstrained in terms of phase and amplitude. For ABF, the beamforming vector $\mathbf{f}_{an}$ is given by
\begin{equation}
\mathbf{f}_{an} = \frac{1}{\sqrt{N_t}} \left[e^{j\varphi_1}, e^{j\varphi_2}, \dots, e^{j\varphi_{N_t}} \right]^T.
\end{equation}
For the fully connected HBF, the beamforming vector $\mathbf{f}_{hb}$ is composed of two components $\mathbf{f}_{hb} = \mathbf{F}_{rf} \mathbf{f}_{bb}$, $\mathbf{f}_{bb} \in \mathbb{C}^{N_t^{RF} \times 1}$ represents the baseband beamforming vector and $\mathbf{F}_{rf} \in \mathbb{C}^{N_t \times N_t^{RF}}$ is the analog beamforming matrix, which can be represented as $\mathbf{F}_{rf} = \frac{1}{\sqrt{N_t}} e^{j\boldsymbol{\Phi}_{rf}}$, here $\boldsymbol{\Phi}_{rf} \in \mathbb{R}^{N_t \times N_t^{RF}}$ is the phase matrix of the analog beamforming. Let the transmitted symbol be $x$ with the power constraint $\mathbb{E}{\left[|x|^2\right]}=1$, the received symbol is
\begin{equation}
 y = \sqrt{\rho} \mathbf{h}^H \mathbf{f} x + n,
\end{equation}
where $ \mathbf{h} \in \mathbb{C}^{N_t \times 1} $ is the channel vector between the BS and UE, $ \rho $ is the transmit power and $ n $ is the complex Gaussian noise in the channel with variance $\sigma^2$. The spectral efficiency is given by
\begin{equation}
\label{eq:SE}
 R = \log_2 \left( 1 + \frac{\rho}{\sigma^2} \mathbf{h}^H \mathbf{f} \mathbf{f}^H \mathbf{h} \right).
\end{equation}

\begin{figure}[t]
\centerline{\includegraphics[width=0.25\textwidth]{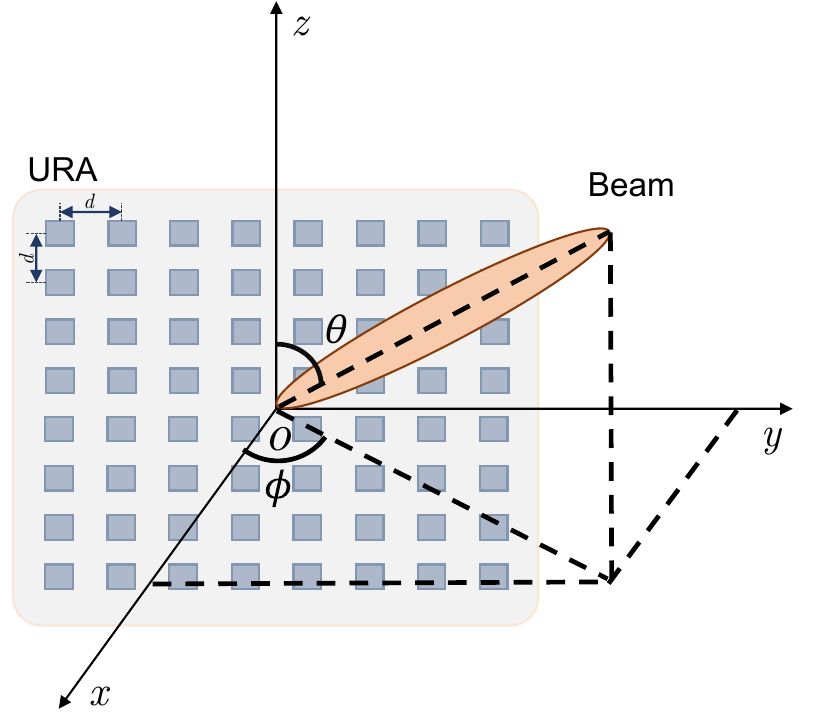}}
\caption{Uniform rectangular array model with a far-field directional beam.}
\label{fig-URA}
\end{figure}

The URA is located in the $yoz$ plane as shown in Fig. \ref{fig-URA}. Given a beamforming vector $\mathbf{f}$, the two-dimensional far-field beam pattern $\mathbf{X} \in \mathbb{R}^{H \times W}$ is computed via the array factor \footnote{Potential mutual coupling among antennas\cite{COUP} is not considered herein and is deferred to future work.}
\begin{equation}
\label{eq:array_factor}
\begin{split}
[\mathbf{X}]_{i,j} &= \log\left|\sum_{m=0}^{N_y-1} \sum_{n=0}^{N_z-1} \mathrm{f}_l e^{jkd\left( m \sin\theta_i \sin\phi_j + n \cos\theta_i \right)}\right|,\\
l &=  m N_z + n + 1,
\end{split}
\end{equation}
where $ k $ is the wavenumber related to the wavelength $ \lambda $ by $ k \lambda = 2\pi $,   $N_z$ and $N_y$ represent the number of antennas in the vertical and horizontal dimensions, $ d $ is the antenna spacing, $ \theta_i $ and $ \phi_j $ denote the $i$-th zenith and $j$-th azimuth angles, respectively, $\mathrm{f}_l$ represents the $l$-th element of $\mathbf{f}$.

\subsection{Problem Formulation}
\begin{figure*}[t]
    \centering
    \includegraphics[width=0.8\textwidth]{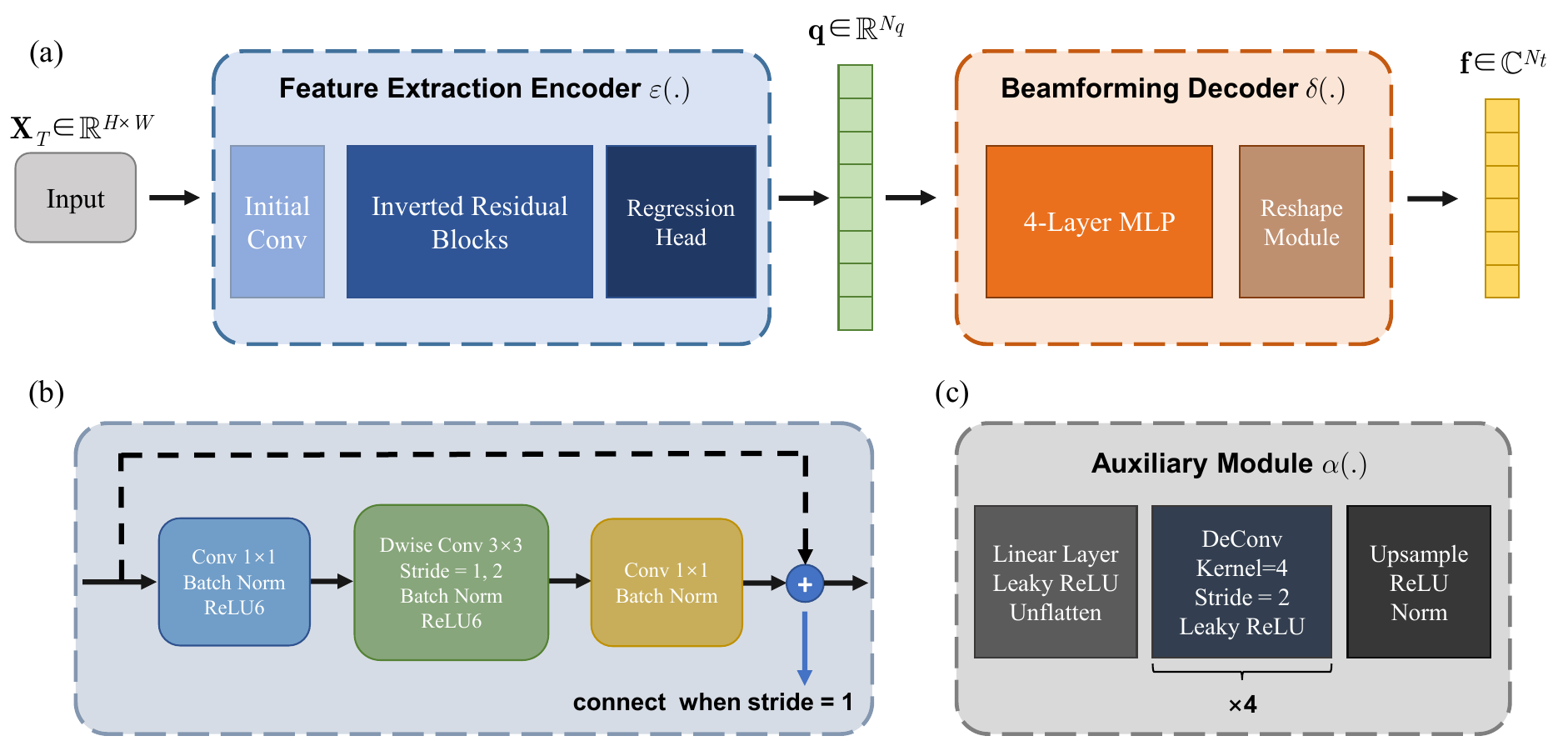}
    \caption{(a) The architecture of the proposed DNN. (b) The structure of inverted residual block in the encoder. (c) Auxiliary module architecture used in the online training process of encoder.}
    \label{fig:arch}
\end{figure*}
We aim to determine the optimal beamforming vector $ \mathbf{f^{*}} $ under the given beamforming architecture such that the spectral efficiency in the considered channel is maximized, which can be formulated as
\begin{equation}
 \mathbf{f^{*}} = \arg\max_{\mathbf{f} \in \mathcal{F}} R
\end{equation}
where $ \mathcal{F} $ represents the solution space of the corresponding beamforming architecture, here we consider the beamforming design under all three architectures mentioned above.

In practice, obtaining complete CSI is challenging due to accuracy limitations and excessive time consumption. Thus we focus on designing the beamforming matrix under partial CSI conditions. We propose a beamforming method based on a target beam pattern, aiming to determine the optimal beamforming vector by minimizing the discrepancy between the obtained beam pattern and the target beam pattern. 

Let $ \mathbf{X}_T \in \mathbb{R}^{H \times W} $ denote the target beam pattern derived from the beam information, where $ W $ and $ H $ represent the number of azimuth and zenith angles, respectively. The beam pattern corresponding to a beamforming vector $ \mathbf{f} $ is denoted as $ \mathbf{X}_f $, which is computed based on the array factor formula (\ref{eq:array_factor}). The optimization problem can then be formulated as
\begin{equation}
 \mathbf{f^{*}} = \arg\min_{\mathbf{f} \in \mathcal{F}} \mathcal{L}(\mathbf{X}_T, \mathbf{X}_f)
\end{equation}
where the function $ \mathcal{L}(\mathbf{X}, \mathbf{Y})$ measures the discrepancy between $ \mathbf{X} $ and $ \mathbf{Y} $. The design of an appropriate function $\mathcal{L}$ to facilitate effective beamforming will be discussed in the next section.

\section{Proposed Method: Deep Learning-Based Beamforming Design With Target Beam Pattern}

We propose a DNN for beamforming design based on target beam patterns. The proposed DNN takes the target beam pattern as input and outputs the corresponding beamforming vector. In this section, we present the architecture, training process, and the design of the loss function of the DNN.

\subsection{Architecture of the Deep Neural Network}
The main challenge in obtaining optimal beamforming vectors from target beam patterns lies in extracting vectors from multidimensional matrices, which can be viewed as a feature extraction problem. While convolutional neural networks (CNNs) are suitable for such tasks, practical wireless communication systems face critical constraints including high power consumption, computational demands, and limited storage capacity. Conventional approaches using CNNs with multiple fully connected layers may suffer from excessive parameterization and training difficulties due to the complexity of various beam patterns and the constraints imposed by different beamforming architectures. To address these challenges, a lightweight encoder-decoder structured DNN is proposed as illustrated in Fig. \ref{fig:arch}. Compared to computationally intensive and data-hungry Vision Transformers and deep CNNs, the encoder-decoder offers a compact and efficient solution, better capturing local structure in beam patterns with reduced complexity. 

The proposed DNN consists of an encoder module and a decoder module. The encoder module extracts essential feature vectors from complex beam patterns, subsequently, the decoder reconstructs beamforming vectors from these encoded features. The overall process can be formulated as
\begin{equation}
\mathbf{f} = \delta\left(\varepsilon\left(\mathbf{X}_{T}, \boldsymbol{\Theta}_e\right), \boldsymbol{\Theta}_d\right),
\end{equation}
where $\mathbf{X}_{T}$ represents the input beam pattern, $\mathbf{f}$ is the output beamforming vector, $\varepsilon(\cdot)$ and $\delta(\cdot)$ denote the encoder and decoder functions, respectively, and $\boldsymbol{\Theta}_e$ and $\boldsymbol{\Theta}_d$ are the trainable parameters of the corresponding modules. 

For the encoder design, we choose MobileNetV2\cite{MobileNet} as the backbone to balance feature extraction capability with computational efficiency. The lightweight architecture encompasses two key innovations: depth-wise separable convolutions and inverted residual blocks. Depth-wise separable convolutions decouple spatial and channel-wise feature processing by first applying single-channel spatial convolutions followed by $1\times1$ point-wise convolutions, significantly reducing computational complexity. Inverted residual blocks enhance feature representation through an expansion-compression strategy: channel dimensions are first expanded via $1\times1$ convolutions, followed by depth-wise convolutions for spatial feature extraction, and finally compressed back to lower dimensions. 

The encoder implementation details are as follows: The input beam pattern undergoes initial convolution through a $3\times3$ convolution layer with batch normalization and ReLU6 activation. Seven inverted residual blocks are then stacked, with configurations progressively increasing channel dimensions and downsampling spatial resolutions. For instance, the first block uses a $6\times$ expansion ratio, 16 output channels, and stride 1. All convolutional layers utilize ReLU6 activation to constrain output ranges for low-precision compatibility. Finally, the regression head, which consists of a $1\times1$ convolution, global average pooling, and a fully connected layer, is used to produce a fixed-length feature vector.

The decoder generates beamforming vectors from encoded features using a multilayer perceptron (MLP) with four hidden layers. The first three layers employ ReLU activation, while the output layer directly generates beamforming parameters. Architectural constraints are addressed through specialized output configurations: 
\subsubsection{Digital beamforming}
The MLP outputs real and imaginary components of the beamforming vector $\mathfrak{R}\{\mathbf{f}_{fd}\}$ and $\mathfrak{I}\{\mathbf{f}_{fd}\}$, yielding $2N_t$ dimensions.
\subsubsection{Hybrid beamforming} MLP outputs parameters that include analog phase shifts $\mathrm{vec}(\boldsymbol{\Phi}_{rf})$ and baseband beamforming components $\mathfrak{R}\{\mathrm{vec}(\mathbf{f}_{bb})\}$ and $\mathfrak{I}\{\mathrm{vec}(\mathbf{f}_{bb})\}$, totaling $(N_t + 2)N_t^{RF}$ dimensions.
\subsubsection{Analog beamforming}Initial experiments revealed suboptimal performance when directly predicting phase components. Instead, we normalize the magnitude of the hybrid beamforming vector and adopt it as the analog beamforming vector.

The MLP output vector $\mathbf{v}$ is reshaped according to these configurations to obtain the final beamforming vector \(\mathbf{f}\). This unified framework enables flexible adaptation to diverse beamforming architectures while maintaining computational efficiency.

\subsection{Training Process and Loss Function Design}
Most existing DL-based beamforming methods utilize offline training. However, in practical applications, wireless communication systems may need to operate in diverse environments, while entirely offline training approaches require retraining distinct networks for each configuration change, incurring huge computational overhead. Additionally, preliminary simulation results reveal that simultaneously training the entire DNN tends to converge to suboptimal solutions. To address these issues, we decompose the training process into two stages.
\subsubsection{Offline pre-training of the encoder}
In the first stage, we train the encoder $\varepsilon(\cdot)$ to capture the intricate features of beam patterns and represent them as feature vectors $\mathbf{q}$, while keeping the decoder untrained. The objective is to ensure that $\mathbf{q}$ retains sufficient information to reconstruct the original beamforming vectors. To efficiently train $\varepsilon(\cdot)$, we introduce an auxiliary module $\alpha(\cdot)$, composed of deconvolutional neural networks, to reconstruct beam patterns from $\mathbf{q}$. During offline pre-training, the input beam pattern $\mathbf{X}$ is processed by $\varepsilon(\cdot)$ to generate $\mathbf{q}$. This vector is then fed into $\alpha(\cdot)$ to produce the reconstructed beam pattern $\hat{\mathbf{X}}$. The training phase aims to minimize the discrepancy between the original beam pattern $\mathbf{X}$ and the reconstructed $\hat{\mathbf{X}}$, formulated as  
\begin{equation}
\mathbf{q} = \varepsilon(\mathbf{X}, \boldsymbol{\Theta}_e), \quad \hat{\mathbf{X}} = \alpha(\mathbf{q}, \boldsymbol{\Theta}_a).
\end{equation}
where $\boldsymbol{\Theta}_a$ denotes the trainable parameters of the auxiliary module. Through joint training of $\varepsilon(\cdot)$ and $\alpha(\cdot)$, $\varepsilon(\cdot)$ acquires stable and representative feature extraction capabilities. The loss function for this stage $\mathcal{L}_1$ is designed as the mean squared error between the input and reconstructed beam patterns.
\begin{equation}
\mathcal{L}_1 = \frac{1}{N_1HW} \sum_{k=1}^{N_1} \sum_{i=1}^H \sum_{j=1}^W \left| [\mathbf{X}_k]_{i,j} - [\hat{\mathbf{X}}_k]_{i,j} \right|^2,
\end{equation}
where $N_1$ represents the number of training samples, $W$ and $H$ denote the width and height of the beam pattern, corresponding to the number of angle samples in the azimuth and zenith dimensions, respectively.

Regarding the dataset for offline pre-training, since this stage employs unsupervised learning, the dataset only requires normalized beam pattern magnitudes. To enable the encoder to learn diverse beam pattern features, the dataset must exhibit sufficient randomness. We generate a large number of random digital beamforming vectors and compute their corresponding beam patterns by (\ref{eq:array_factor}), with detailed parameter configurations in Section IV.
\subsubsection{Online training of the decoder}
After acquiring the pre-trained $\varepsilon(\cdot)$ through offline pre-training, the second stage concentrates on optimizing $\delta(\cdot)$. The target beam pattern $\mathbf{X}_T$ is encoded into a feature vector $\hat{\mathbf{q}}$ via pre-trained $\varepsilon(\cdot)$, which is subsequently fed into $\delta(\cdot)$ to generate the beamforming vector $\mathbf{f}$. The corresponding beam pattern $\mathbf{X}_f$ is then computed. The process can be expressed as  
\begin{equation}
\hat{\mathbf{q}} = \varepsilon(\mathbf{X}_T, \boldsymbol{\Theta}_e^*), \quad \mathbf{f} = \delta(\hat{\mathbf{q}}, \boldsymbol{\Theta}_d).
\end{equation}
where $\boldsymbol{\Theta}_e^*$ represents the fixed parameters of the pre-trained $\varepsilon(\cdot)$. As discussed in Section II, the design of an appropriate loss function $\mathcal{L}_2$ is critical for beamforming optimization. To meet practical requirements, $\mathbf{X}_f$ should exhibit three key characteristics: close alignment with $\mathbf{X}_T$ in the main lobe region\footnote{The main lobe region is defined as the areas with normalized amplitude exceeding -10 dB in the dB-scale beam pattern.}, suppressed side lobe\footnote{The side lobe region corresponds to the areas with amplitude below -20 dB.} levels below those of $\mathbf{X}_T$, and similarity in the moderate region (between -20 dB and -10 dB). Accordingly, the composite loss function is formulated as 
\begin{equation}
\mathcal{L}_2 = \mathcal{L}_{ml} + \mathcal{L}_{sl} + \mathcal{L}_{md},
\end{equation}
where the individual components are defined as
\begin{subequations}
\label{eq:loss_components}
\begin{align}
\mathcal{L}_{ml} &= \frac{1}{N_2 N_{ml}}\sum_{k=1}^{N_2} \sum_{(i,j)\in \mathcal{M}} \left| [\mathbf{X}_{T,k}]_{i,j} - [\mathbf{X}_{f,k}]_{i,j} \right|^2,  \label{eq:Lml} \\
\mathcal{L}_{sl} &= \frac{1}{N_2 N_{sl}} \sum_{k=1}^{N_2} \sum_{(i,j)\in \mathcal{S}} \left| \max\left(0, [\mathbf{X}_{f,k}]_{i,j} - [\mathbf{X}_{T,k}]_{i,j}\right) \right|^2,  \label{eq:Lsl} \\
\mathcal{L}_{md} &= \frac{1}{N_2 N_{md}} \sum_{k=1}^{N_2} \sum_{(i,j)\in \mathcal{T}} \left| [\mathbf{X}_{T,k}]_{i,j} - [\mathbf{X}_{f,k}]_{i,j} \right|^2,  \label{eq:Lmd}
\end{align}
\end{subequations}
here $\mathcal{M}$, $\mathcal{S}$ and $\mathcal{T}$ denote the main lobe, side lobe and moderate regions, respectively. $N_{ml}$, $N_{sl}$ and $N_{md}$ represent the dimensions of these regions, while $N_2$ is the number of training samples in this stage. The max operator in $\mathcal{L}_{sl}$ ensures penalization only when $\mathbf{X}_f$ exceeds the target side lobe level.

\section{Numerical Results}

\subsection{Simulation Setup}

\begin{figure}[t]
\centerline{\includegraphics[width=0.4\textwidth]{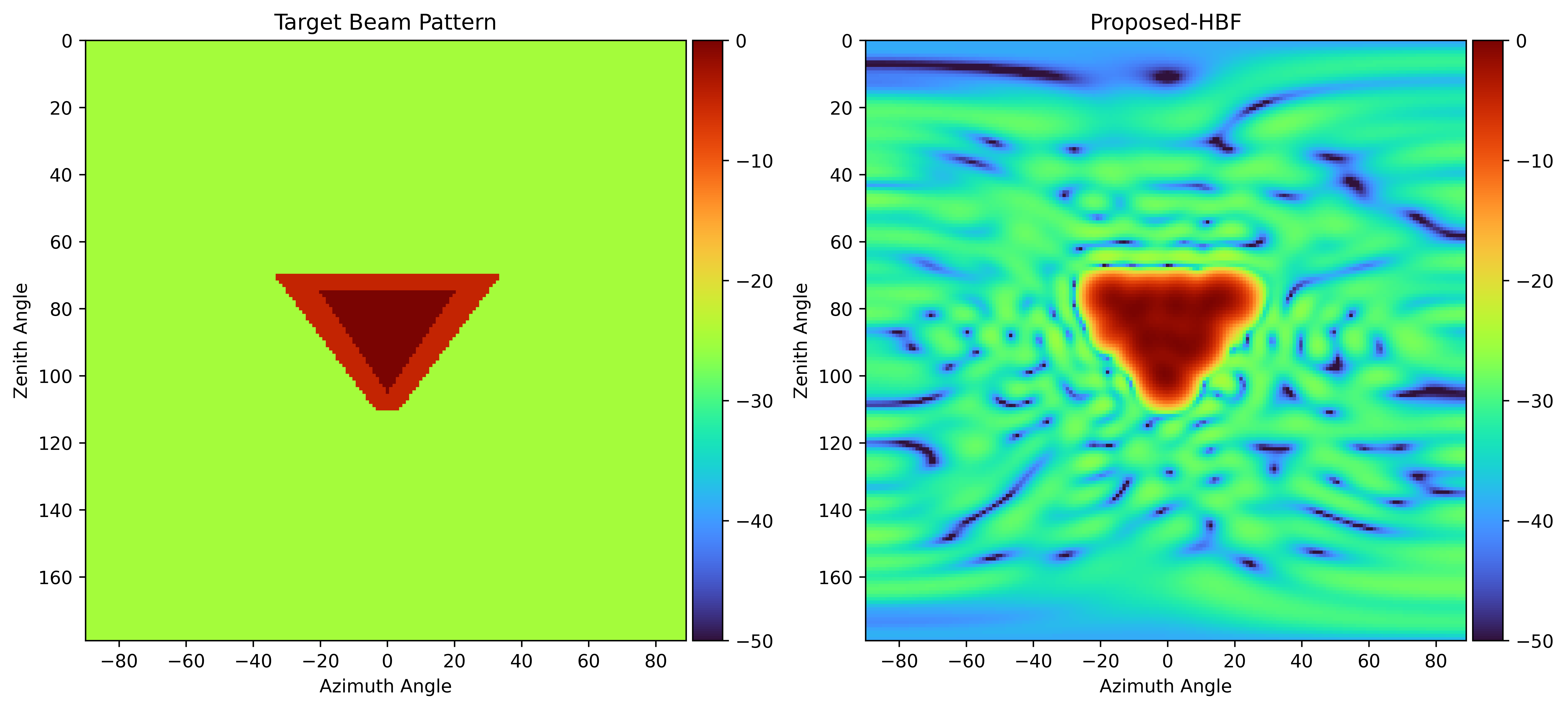}}
\caption{Target beam pattern (left) with a triangular main lobe shape which has a base length of $40\degree$ and a height of $30\degree$, and the beam pattern predicted by the proposed method under HBF (right).}
\label{fig-triangle}
\end{figure}

\begin{figure}[t]
\centerline{\includegraphics[width=0.42\textwidth]{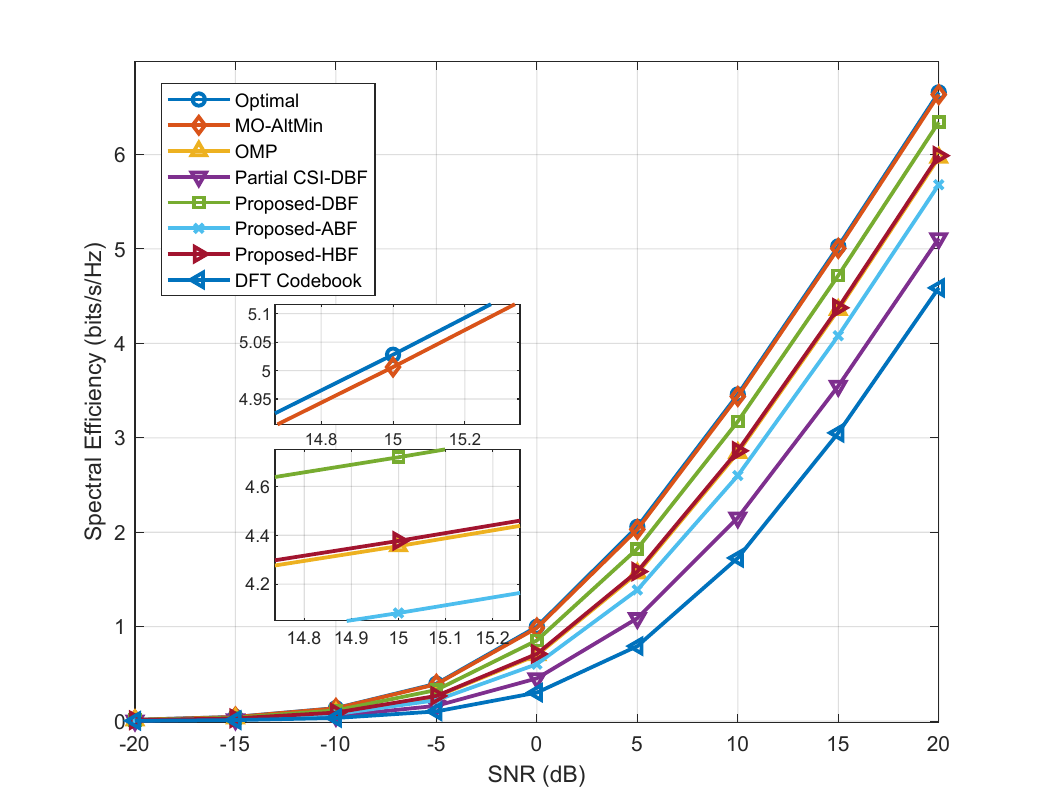}}
\caption{Spectral efficiency achieved by different beamforming methods under NYUSIM-generated mmWave channels.}
\label{fig-SE}
\end{figure}

\begin{figure*}[t]
    \centering
    \includegraphics[width=0.7\textwidth]{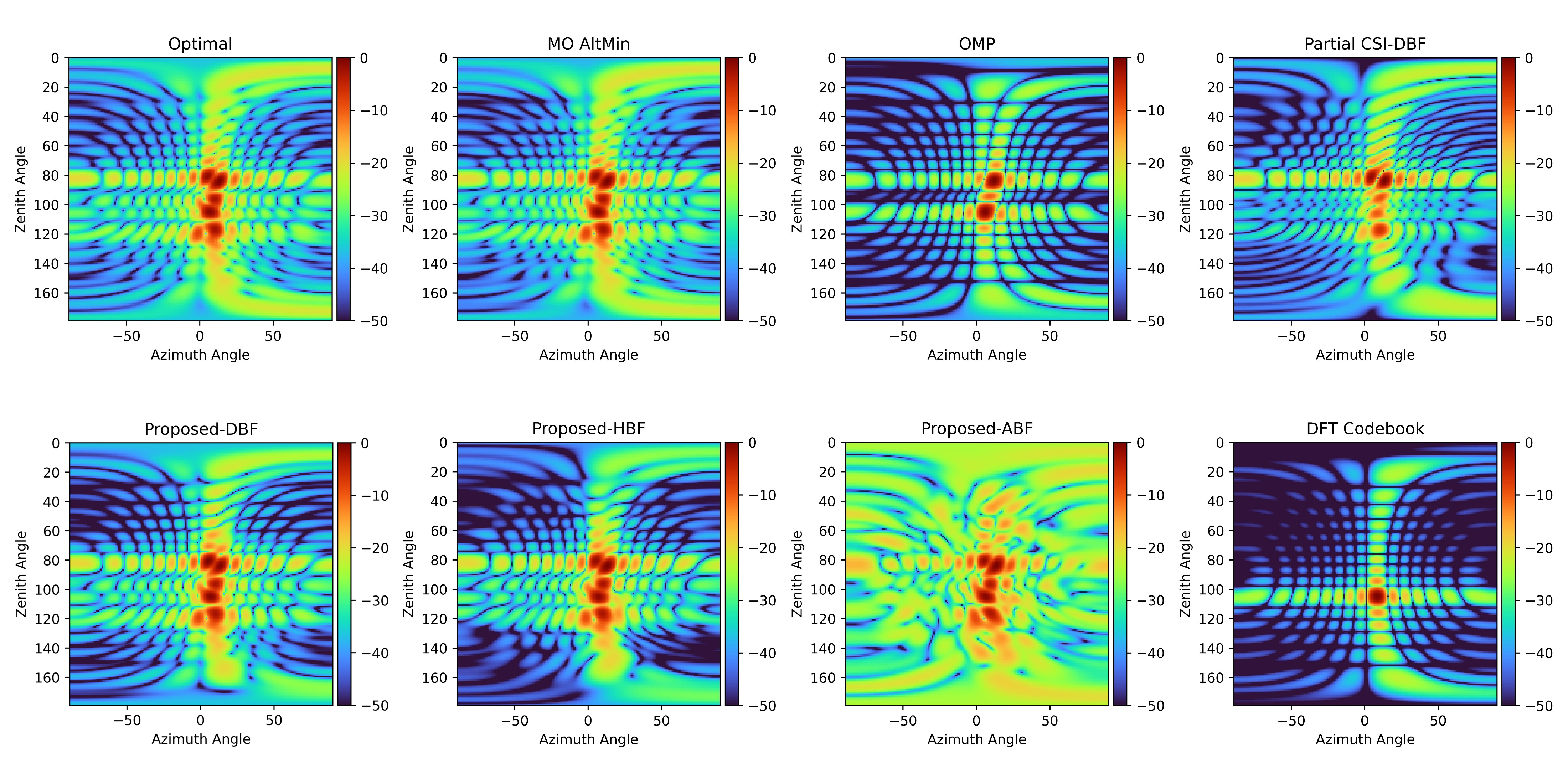}
    \caption{Normalized beam patterns generated by different beamforming methods for one non-line-of-sight mmWave channel vector from online training dataset.}
    \label{fig:patterns}
\end{figure*}
We assume the BS is equipped with a URA of $N_y = N_z = 16$ elements and UE is equipped with a single antenna. The number of RF chains for HBF $N_{t}^{RF} = 2$. The operating frequency of the communication system is set to $28$ GHz and the antenna spacing $d$ is set to $\frac{\lambda}{2}$. For the size of the beam pattern, the range of the azimuth angle is set to {-89\degree} to 90\degree, the range of the zenith angle is set to {1\degree} to 180\degree, and the angle sampling interval is 1\degree, which means that both $W$ and $H$ are 180.

In the proposed DNN, the dimension of feature vector $N_q = 1024$, and the dimensions of each layer in decoder are $1024$, $2048$, $1024$ and $N_v$, respectively, where $N_v$ represents the dimension of MLP output vector, which is $2N_t$ or $(N_t+2)N_t^{RF}$ for digital or hybrid beamforming. The size of the dataset used for pre-training is $10000$, which is randomly generated. $90\%$ of the dataset are used for train and the remaining are validation samples. The learning rate of pre-training is $10^{-5}$ using the Adam optimizer, the total pre-training epoch is $2000$ and the batch size is $1024$. As for online training, the channel data is generated from the channel simulator NYUSIM\cite{NYUSIM}, which can generate channels with high fidelity. The target beam pattern dataset is derived from the optimal DBF vector of the channel data. In this paper, we consider a system with limited channel information and a relatively simple working mode. In the online deployment, only a few target beam patterns are needed for the service scenario. Therefore, the size of online training pattern samples is $10$. The online learning rate and training epoch are $10^{-3}$ and $500$ respectively. The experiments are performed on an AMD Ryzen 5 7500F CPU and 8 48GB NVIDIA A40 GPUs with PyTorch of CUDA 11.4.

\subsection{Performance Evaluation}
First, we evaluate the performance of complex pattern synthesis in Fig. \ref{fig-triangle}. Here the target beam pattern is configured with an isosceles triangular main lobe and a side lobe level of -25 dB. We take HBF as an example, and the result shows that the predicted beam pattern meets the requirements of the target beam pattern, in the main lobe region the amplitude fluctuation is within 2 dB and the amplitude in the side lobe region is generally smaller than the target. Total online training time is about 52s, and using genetic algorithm\cite{GA} takes about 600s.

Furthermore, we explore the beamforming performance on NYUSIM-generated millimeter-wave (mmWave) channels in Fig. \ref{fig-SE}. Here we choose optimal DBF, MO-AltMin \cite{MO}, OMP\cite{OMP} for HBF which demand full CSI, optimal DBF under partial CSI which means that the BS only knows the amplitude of the complex gain but not its phase\cite{PathLossEst}, and the conventional beamforming scheme based upon DFT codebooks as benchmarks\footnote{Due to differences in array topology and beamforming architecture, methods in \cite{AEArraySyn,AE2D} are not directly applicable to URA-based beamforming.}. Meanwhile, we consider beamforming based on beam patterns containing phase information. This method yields beamforming vectors that are highly similar to the optimal ones. Specifically, let the beam pattern with phase be calculated as $\mathrm{vec}(\mathbf{X}_p)^T = \mathbf{f}^H\mathbf{A}$, where $\mathbf{A} \in \mathbb{C}^{N_t \times WH}$ is constructed from the steering vectors of sampled directions and $\mathbf{f} \in \mathbb{C}^{N_t \times 1}$ is the optimal beamforming vector. When the number of samples satisfies $WH \gg N_t$, the matrix $\mathbf{A} \mathbf{A}^H$ satisfies that $\mathbf{A} \mathbf{A}^H \approx c \cdot \mathbf{I}$, where $c$ is a constant. The least-squares solution is then given by
\begin{equation}
\label{eq:LS}
\hat{\mathbf{f}}^H = \frac{\mathrm{vec}(\mathbf{X}_p)^T}{\mathrm{max}(\mathbf{X}_p)}\mathbf{A}^H(\mathbf{AA}^H)^{-1} \approx  b \cdot\mathbf{f}^H,
\end{equation}
where $b$ is a constant. As a result, with sufficiently dense sampling, this approach accurately recovers the original beamforming vector and achieves nearly identical spectral efficiency to optimal DBF after power normalization. The online training time for digital and hybrid are 61s and 65s while MO-AltMin takes 69s. As illustrated in Fig. \ref{fig-SE}, within the SNR range of -20 to 20 dB, the proposed scheme attains $93\%$ of the optimal fully-digital spectral efficiency, overtakes the method with partial CSI under DBF and significantly outperforms DFT codebook in ABF. For HBF, it marginally exceeds OMP performance but remains inferior to MO-AltMin, yet the latter incurs higher computational complexity due to extensive iterations\cite{MO}. In the considered antenna configuration, HBF has a faster closed-form solution that allows it to achieve the same performance as DBF\cite{CloseForm}, however, it still requires perfect CSI. Notably, the proposed method eliminates the real-time acquisition of full CSI by generating beamforming vectors directly from target beam patterns. In terms of channel estimation overhead, using target beam patterns eliminates the need to estimate the phase of complex path gains. Using the same partial CSI definition in \cite{PathLossEst}, from which a normalized beam pattern can be constructed, could reduce pilot overhead used for channel estimation to a certain extent.

Fig. \ref{fig:patterns} depicts the beam patterns of various beamforming methods in a non-line-of-sight channel in the training samples. It can be seen that the optimal beam pattern, also as the target beam pattern, has multiple complex main lobes with side lobe level below -25 dB. Among these, MO-AltMin and the proposed method perform well in beamforming in the main lobe region, achieving the same number, position, and shape as the target. However, the OMP, DBF under partial CSI and DFT codebook methods can only cover a portion of the main lobe. In the side lobe region, the side lobe levels predicted by the proposed method basically meet the requirement except for ABF. Results show that compared with optimization schemes that require full CSI and codebook methods that require beam training, the proposed method can achieve good beamforming performance with only beam patterns and has advantages in channel estimation overhead.

\section{Conclusion}
In this paper, we have investigated efficient beamforming design with limited CSI, capitalizing on the target beam pattern. A DL-based framework is developed that employs a lightweight encoder-decoder structure to directly map the target beam pattern to the beamforming vector without the acquisition of full CSI. We also propose a two-stage training process in which the encoder and decoder are trained separately, effectively reducing online training time. Simulation results demonstrate that the proposed method can effectively predict the beamforming vectors corresponding to complex beam patterns, achieving spectral efficiency close to optimal under limited CSI for various architectures. Future works may focus on beamforming for multi-user systems and hardware-integrated network optimization to enhance practical deployment.  

\bibliography{references} 
\bibliographystyle{IEEEtran}

\end{document}